\documentclass[aps,pre, onecolumn,showpacs,12pt,preprintnumbers,showkeys]{revtex4}
\usepackage{amsfonts}
\usepackage{graphicx}
\usepackage{dcolumn}
\usepackage{bm}
\include{biblio}

\begin{document}

\title{Effect of Bjerrum pairs on electrostatic properties in an electrolyte solution near charged surfaces: A mean-field approach}

\author{Jun-Sik Sin}
\email{js.sin@ryongnamsan.edu.kp}
\affiliation{Natural Science Center, \textbf{Kim Il Sung} University, Taesong District, Pyongyang, Democratic People's Republic of Korea}

\begin{abstract}
In this paper, we investigate the consequences of ion association, coupled with the considerations of finite size effects and orientational ordering of Bjerrum pairs as well as ions and water molecules, on electric double layer near charged surfaces. Based on the lattice statistical mechanics accounting for finite sizes and dipole moments of ions, Bjerrum pairs and solvent molecules, we consider the formation of Bjerrum pairs and derive the mathematical expressions for Bjerrum pair number density as well as cation/anion number density and water molecule number density. 

We reveal the several significant phenomena. Firstly, it is shown that our approach naturally yields the equilibrium constant for dissociation-association equilibrium between Bjerrum pairs and ions.  Secondly, at low surface charge densities, an increase in the bulk concentration of Bjerrum pairs enhances the permittivity and decreases the differential capacitance. Next, for cases where Bjerrum pairs in an alcohol electrolyte solution have a high value of dipole moment, Bjerrum pair number density increases with decreasing distance from the charged surface, and differential capacitance and permittivity is high compared to ones for the cases with lower values of Bjerrum-pair dipole moments. Finally, we show that the difference in concentration and dipole moment of Bjerrum pairs can lead to some variation in osmotic pressure between two similarly charged surfaces.

\end{abstract}


\maketitle

\section{Introduction}

Thermodynamic properties of electrolyte solutions are determined by interactions among the three species in solution, namely, solvent molecules, anions, and cations. 

In general, strong electrolytes exist in the form of strong acids, strong bases and salts, a typical example being NaCl. When NaCl dissolves in a highly polar solvent such as water, the substance is fully dissociated into cations and anions by ion-dipole interactions with solvent molecules. However, in a solvent with a lower relative permittivity, such as methanol, NaCl is not always completely dissociated into cations and anions, some fraction of ions is paired. i.e. in a low polar solvent, strong electrolytes behave as a weak electrolyte.
On the other hand, a weak electrolyte forms ions by interaction with water molecules, a well-known example being acetic acid. Acetic acid provides solvated protons and acetate ions by interaction with water molecules. Acetic acid molecules are not fully dissociated into ions when the solvent is water. 

In order to estimate the thermodynamic properties of an electrolyte, the dissociation constant of ion pairs formed in an electrolyte must be known \cite%
{book,feng_2019,frydel_2018,bawol_2019,self_2020,aghaie_2007,holovko_2001,zhu_2018,monascal_2018,goodwin_2017,feng_2017}. 	

Bjerrum\cite%
{Bjerrum_1926} was the first to propose the concept of ion pair, assuming that all oppositely charged ions within a certain distance of a central ion are paired.  Although he obtained an estimate of the association constant by using statistical theories, the drawback of the method is that in solutions with low permittivity, the critical distance involved in defining ion pair is unreasonably large.

     To overcome the shortcoming, Fuoss \cite%
{fuoss_1934, fuoss_1958} suggested the theory that the cations in the solution are assumed to be conducting spheres of a certain radius and the anions to be point charges.  However, Fuoss theory has also several difficulties as it does not consider dielectric saturation effects of the solvent and depends on the choice of the effective size for the ions. 

    Fisher and Levin in \cite%
{fisher_1993, levin_1996} extended the Debye-H\"uckel theory \cite%
{debye_1923} by accounting for the existence of Bjerrum pairs, and explained phase separation and criticality in electrolyte solutions, in good agreement with simulation results \cite%
{pa_1992,pa_2002,ro_2000}.

    The authors of \cite%
{roij_2009} studied the effect of Bjerrum pairs on the screening length by means of a modified Poisson- Boltzmann theory accounting for simple association-dissociation equilibrium between free ions and Bjerrum pairs. Consequently, they elucidated that in a lower polar solvent, the screening length can be significantly larger than the Debye length, as reported by Leunissen et al \cite%
{le_2007}

   The authors of \cite%
{adar_2017} developed the nonlinear Poisson-Boltzmann framework based on field-theoretical approach \cite%
{Andelman_2000, levy_2012, levy_2013,Andelman_2014, Andelman_2015}, accounting for free ions and pairs. They demonstrated that as observed in \cite%
{smith_2016}, the screening length can be non-monotonic as a function of the ionic concentration.

    In fact, the formation of ion pairs not only reduces the concentration of free ions participating in screening but also increases relative permittivity of ionic solution by excluding less water dipoles, attributed to decrease of free ions and orientational ordering of ion pairs. 

   However, the previous theories \cite%
{roij_2009, adar_2017} could not take into account the difference in size between ions and solvent molecules, and required too high value of dipole moment of a solvent molecule to fit relative permittivity of aqueous solution. Moreover, attention in the studies has been focused on the properties of bulk electrolyte such as screening length and relative permittivity as a function of salt concentration.

   On the other hand, recent studies \cite%
{ iglic_1996, Borukhov_2000, iglic_2010, iglic_2015, sin_2015, sin_2016, sin_pccp_2016, sin_2017, chu_2007, kornyshev_2007, wen_2012, danil_2012, podgornik_2016} developed the free-energy based mean field theories significantly improved by considering the non-uniform size effects of ions and solvent molecules and orientational ordering of solvent dipoles, but any of them did not present estimate of how Bjerrum pairs affect the electrostatic properties close to charged surfaces. 

   In this paper, we incorporate not only steric effects but also orientational ordering of Bjerrum pair dipoles into mean-field approach based on lattice statistical mechanics. In other words, we extend the previously developed mean-field theories \cite%
{iglic_2010, iglic_2015, sin_2015, sin_2016, sin_pccp_2016, sin_2017} in order to include the formation, size effect and orientational ordering of Bjerrum pairs.  We demonstrate that Bjerrum pairs have a significant effect on electrostatic properties in electrolyte solution near a charged surface as well as electrostatic interaction between two similarly charged surfaces. In particular, we focus on an important role of ion-pair association energy and dipole moment of Bjerrum pairs in determining such electrostatic properties.

\section{Theory}

\begin{figure}
\begin{center}
\includegraphics[width=0.5\textwidth]{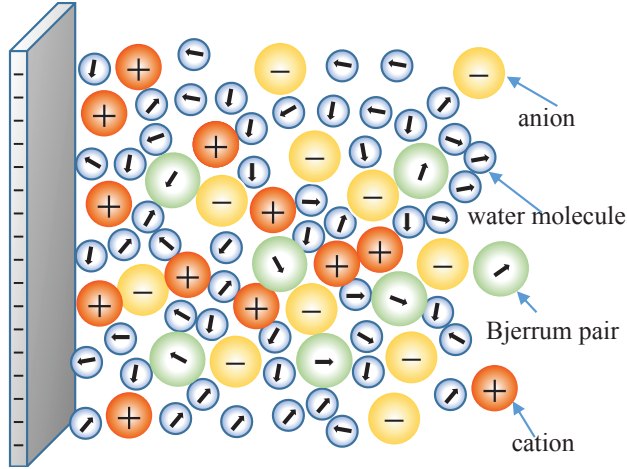}
\caption{(Color online) Schematic of Electric Double Layer formed near a charged surface.  The electrolyte solution contains cations (red circles),  anions (yellow circles), Bjerrum pairs (green circles) and solvent molecules (blue circles).}
\label{fig:1}
\end{center}
\end{figure}
We consider an electrolyte solution with a monovalent electrolyte with cations/anions of elementary charge $e$. The number density for cations, anions, Bjerrum pairs and solvent molecules are denoted as $n_+$, $n_-$ , $n_B$ and $n_w$, respectively, while bulk number densities of them $n_{+b}$, $n_{-b}$, $n_{Bb}$, $n_{wb}$.

The anions and cations have the same bulk concentrations $n_b$, satisfying electro-neutrality. The solvent molecules are modeled as dipoles having permanent dipole moment of ${\bf p}_{w}$. We further assume that some fraction of the cations and anions form Bjerrum pairs that are modeled as dipoles with moment ${\bf{p}}_{B}=Le$, where length $L$ denotes a mean separation between the paired ions. The bulk number density of free ions is $n_{+b}$, the number density of Bjerrum pairs being $n_{Bb}$, satisfying  $n_{+b}+n_{Bb}=n_b$, where $n_b$ is associated with the bulk ion concentration as $n_b=c_b\times N_A\times 1000$ and $N_A$ is the Avogadro number.

Here, $-J$ is the ion-pair association energy which accounts for the electrostatic attraction and short-range interactions.  We restrict ourselves to positive $J$ values, $J>0$. Note that negative infinity($J=-\infty$) denotes the case without Bjerrum pairs. 

We assign the effective volume of the solvent molecules, cations, anions, and Bjerrum pairs as $V_w$, $V_+$, $V_-$ and $V_B$, respectively. 
We assume that all lattice sites are occupied by cations, anions, Bjerrum pairs and solvent molecules.
The total free energy density of the system consisting of the electrode and an aqueous electrolyte can be written as follows.
 \begin{eqnarray}
f =  - \frac{{\varepsilon _0 \left( {\nabla \psi } \right)^2 }}{2} + e\left( {n_ +   - n_ -  } \right)\psi  + \left\langle {\rho _w p_w E\cos \omega } \right\rangle  - ST - \mu _ +  n_ +   - \mu _ -  n_ -  - \mu _w n_w \nonumber \\ + \left( {\left\langle {\rho _B p_B E\cos \omega } \right\rangle  - \mu _B n_B } \right),
\label{eq:1}
\end{eqnarray}
where $\left\langle {g\left( \omega  \right)} \right\rangle  = \int_{}^{} {g\left( \omega  \right)2\pi \sin \left( \omega  \right)d\omega } /4\pi$ in which $\omega$ is the angle between the dipole moment vector and the normal to the charged surface. $\varepsilon_0$ is the permittivity of the vaccum and $\bf{E}$ the electric field strength, while $p_w  = \left| {{\bf{p}}_w } \right|,p_B  = \left| {{\bf{p}}_B } \right|,E = \left| {\bf{E}} \right|,
n_B  = \left\langle {\rho _B \left( \omega  \right)} \right\rangle, n_w  = \left\langle {\rho _w \left( \omega  \right)} \right\rangle$. 

In Eq.(\ref{eq:1}), the first term represents the self-energy of electric field, the second one is the electrostatic potential of anions and cations, the third is the electrostatic potential of solvent dipoles, the fifth one accounts for entropy contribution to the free energy, and the next three terms mean the chemical potential of cations, anions and solvent molecules and the final terms corresponds to the free energy related to Bjerrum pairs. $T$ is the absolute temperature and $k_B$ is the Botlzmann constant.
The number of arrangement can be written as the following expression
 \begin{equation}
\small
W = \frac{{\left( {n_ +   + n_ -   + n_w  + n_B } \right)!}}{{n_ +  !n_ -  !n_w !n_B !}} \cdot \frac{{n_B !}}{{\rho _B \left( {\omega _1 } \right)!\rho _B \left( {\omega _2 } \right)!...\rho _B \left( {\omega _m } \right)!}} \cdot \frac{{n_w !}}{{\rho _w \left( {\omega _1 } \right)!\rho _w \left( {\omega _2 } \right)!...\rho _w \left( {\omega _m } \right)!}},
\label{eq:2}
\end{equation}
which can be used only for low salt concentrations.
Here we consider $n_w = \sum\limits_{i = 1}^m {\rho_{w} \left( {\omega _i } \right)}$ and $n_B = \sum\limits_{i = 1}^m {\rho_{B} \left( {\omega _i } \right)}$. 

However, Eq. (\ref{eq:2}) can be applied to lower ionic concentrations, satisfying volume fraction $\phi<0.1$.

Although the above formula is suitable for the cases where either all ions and dipoles have an identical size or bulk ion cocentration is low, previous studies\cite%
{sin_2015, sin_2016}  show that the formula remains some reasonable also for medium salt concentration.
 \begin{eqnarray}
\small
S = k_B \ln W = k_B \left[ \left( {n_ +   + n_ -   + n_B  + n_w } \right)\ln \left( {n_ +   + n_ -   + n_B  + n_w } \right) - n_ +  \ln n_ +   - n_ -  \ln n_ - \right] \nonumber \\ + k_B\left[-\left\langle {\rho _w \ln \frac{{\rho _w }}{{n_w }}} \right\rangle  - \left\langle {\rho _B \ln \frac{{\rho _B }}{{n_B }}} \right\rangle \right].
\label{eq:3}
\end{eqnarray}
It is assumed that in the present physical system, the incompressibility condition is always satisfied.
\begin{equation}
1 = n_ +  V_ +   + n_ -  V_ +   + n_B V_B  + n_w V_w. 
\label{eq:4}
\end{equation}
The number densities of cations, anions, Bjerrum pairs and solvent dipoles and electrostatic potential are obtained from the fact that the free energy of the system has an extreme value at thermodynamic equilibrium of the whole system, satisfying the incompressibility condition.
The Lagrangian of the system is expressed as follows
\begin{equation}
L = \int_{}^{} {fdx}  - \int_{}^{} {\lambda \left( x \right)\left( {1 - n_ +  V - n_ -  V - n_w V_w  - n_B V_B } \right)},
\label{eq:5}
\end{equation}
where $\lambda \left(x\right)$ is a  local Lagrange parameter. 
The Euler-Lagrange equation with respect to the cation number density yields the following equation.
\begin{equation}
\frac{{\delta L}}{{\delta n_ +  }} = \frac{{\partial L}}{{\partial n_ +  }} = e\psi  - \mu _ +   + k_B T\ln \left( {n_ +  /\left( {n_ +   + n_ -   + n_w  + n_B } \right)} \right) + \lambda V_ +   = 0.
\label{eq:6}
\end{equation}
As $x$ gets far away from the charged surface, the following equations are satisfied.
\begin{equation}
 \psi  = 0, \lambda  = \lambda _b, n_ +  = n_{ + b}, n_ -   = n_{ - b}, n_B  = n_{Bb}. 
\label{eq:7}
\end{equation}
Considering the above facts and substituting  Eq. (\ref{eq:7}) into Eq. (\ref{eq:6}) results in the following expression for chemical potential of cations.
\begin{equation}
\mu _ +   = k_B T\ln \frac{{n_{ + b} }}{{\left( {n_{ + b}  + n_{ - b}  + n_{Bb}  + n_{wb} } \right)}} + \lambda _b V_ + . 
\label{eq:8}
\end{equation}
In the same way, chemical potentials of the anions, solvent molecules and Bjerrum pairs are as follows
\begin{equation}
\mu _ -   = k_B T\ln \frac{{n_{ - b} }}{{\left( {n_{ + b}  + n_{ - b}  + n_{Bb}  + n_{wb} } \right)}} + \lambda _b V_ -.  
\label{eq:9}
\end{equation}
\begin{equation}
\mu _w  = k_B T\ln \frac{{n_{wb} }}{{\left( {n_{ + b}  + n_{ - b}  + n_{Bb}  + n_{wb} } \right)}} + \lambda _b V_w. 
\label{eq:10}
\end{equation}
\begin{equation}
\mu _B  = k_B T\ln \frac{{n_{Bb} }}{{\left( {n_{ + b}  + n_{ - b}  + n_{Bb}  + n_{wb} } \right)}} + \lambda _b V_B. 
\label{eq:11}
\end{equation}
From the fact that a Bjerrum pair is formed by combination of a cation and an anion, we now can recognize the following relation between chemical potentials of Bjerrum pairs, cations and anions
\begin{equation}
\mu _B  = \mu _ +   + \mu _ -   + J.
\label{eq:12}
\end{equation}
With the help of Eqs. (\ref{eq:8},\ref{eq:9},\ref{eq:11}), the relation between bulk ion number density and bulk Bjerrum pair number density can be established straightforwardly. 
\begin{eqnarray}
 k_B T\ln \frac{{n_{Bb} }}{{\left( {n_{ + b}  + n_{ - b}  + n_{Bb}  + n_{wb} } \right)}} + \lambda _b V_B  - J - \left( {k_B T\ln \frac{{n_{ + b} }}{{\left( {n_{ + b}  + n_{ - b}  + n_{Bb}  + n_{wb} } \right)}} + \lambda _b V_ +  } \right) \nonumber \\ 
  - \left( {k_B T\ln \frac{{n_{ - b} }}{{\left( {n_{ + b}  + n_{ - b}  + n_{Bb}  + n_{wb} } \right)}} + \lambda _b V_ -  } \right) = 0 . 
\label{eq:13} 
\end{eqnarray}
After further manipulation, we obtain the following expression for the equilibrium constant.
\begin{eqnarray}
\small
K=\frac{{\left( {n_b  - n_{Bb} } \right)^2 }}{{n_{Bb} }}=  \frac{n_{ +b}n_{-b}}{{n_{Bb} }} = \left( {n_{ + b}  + n_{ - b}  + n_{Bb}  + n_{wb} } \right)\exp \left( { - \lambda _b \left( {V_ +   + V_ -   - V_B } \right)/k_B T} \right)\exp \left( { - J/k_B T} \right). 
\label{eq:14}
\end{eqnarray}
Like in previous studies \cite%
{adar_2017}, Eq. (\ref{eq:14}) means that the number density of Bjerrum pairs increases with magnitude of ion-pair association energy. The formula also says that the larger the difference between the sum of volumes of a cation and an anion and the volume of a Bjerrum pair, the more Bjerrum pairs are formed.

Inserting Eq. (\ref{eq:8}) into Eq.(\ref{eq:6}) and after further manipulations, we obtain the following equations  

\begin{equation}
\frac{{n_ +  }}{{n_{ + b} }}\frac{{n_{ + b}  + n_{ - b}  + n_{Bb}  + n_{wb} }}{{n_ +   + n_ -   + n_B  + n_w }} = \exp \left( { - \left( {hV_ +   + e\psi } \right)/k_B T} \right),
\label{eq:15}
\end{equation}
\begin{equation}
\frac{{n_ -  }}{{n_{ - b} }}\frac{{n_{ + b}  + n_{ - b}  + n_{Bb}  + n_{wb} }}{{n_ +   + n_ -   + n_B  + n_w }} = \exp \left( { - \left( {hV_ -   - e\psi } \right)/k_B T} \right),
\label{eq:16}
\end{equation}
\begin{equation}
\frac{{n_B }}{{n_{Bb} }}\frac{{n_{ + b}  + n_{ - b}  + n_{Bb}  + n_{wb} }}{{n_ +   + n_ -   + n_B  + n_w }} = \exp \left( { - hV_B /k_B T} \right)\frac{{\sinh \left( {p_B E/\left(k_BT\right)} \right)}}{{\left( {p_B E/\left(k_BT\right)} \right)}},
\label{eq:17}
\end{equation}
\begin{equation}
\frac{{n_w }}{{n_{wb} }}\frac{{n_{ + b}  + n_{ - b}  + n_{Bb}  + n_{wb} }}{{n_ +   + n_ -   + n_B  + n_w }} = \exp \left( { - hV_w /k_B T} \right)\frac{{\sinh \left( {p_w  E/\left(k_BT\right)} \right)}}{{p_w  E /\left(k_BT\right)}},
\label{eq:18}
\end{equation}
where $h=\lambda-\lambda_b$.

Multiplying Eqs.(\ref{eq:15},\ref{eq:16},\ref{eq:17},\ref{eq:18}) by $V_ +  ,V_ -  ,V_B ,V_w$, respectively, and adding the each equations, we obtain
\begin{eqnarray}
 \frac{{n_{ + b}  + n_{ - b}  + n_{Bb}  + n_{wb} }}{{n_ +   + n_ -   + n_B  + n_w }} = n_{ + b} V_ +  e^{\left( { - hV_ +   - e\psi } \right)/k_B T}  + n_{ - b} V_ -  e^{\left( { - hV_ -   + e\psi } \right)/k_B T} \nonumber \\ 
  + n_{wb} V_w e^{ - hV_w /k_B T} \frac{{\sinh \left( {p_w E/\left(k_BT\right)} \right)}}{{p_w E/\left(k_BT\right)}} + n_{Bb} V_B e^{ - hV_B /k_B T} \frac{{\sinh \left( {p_B E/\left(k_BT\right)} \right)}}{{p_B E/\left(k_BT\right)}}.  
\label{eq:19}
\end{eqnarray}
Substituting Eq. (\ref{eq:19}) in Eq. (\ref{eq:15},\ref{eq:16},\ref{eq:17},\ref{eq:18}) results in the expressions for number densities of cations, anions, Bjerrum pairs and solvent dipoles:
\begin{equation}
n_ +   = \frac{{n_{ + b} e^{\left( { - hV_ +   - e\psi } \right)/k_B T} }}{D},
\label{eq:20}
\end{equation}
\begin{equation}
n_ -   = \frac{{n_{ - b} e^{\left( { - hV_ -   + e\psi } \right)/k_B T} }}{D},
\label{eq:21}
\end{equation}
\begin{equation}
n_w  = \frac{{n_{wb} e^{\left( { - hV_w /k_B T} \right)} \frac{{\sinh \left( {p_w E/\left(k_BT\right)} \right)}}{{\left( {p_w E/\left(k_BT\right)} \right)}}}}{D},
\label{eq:22}
\end{equation}
\begin{equation}
n_B  = \frac{{n_{Bb} e^{\left( { - hV_B /k_B T} \right)} \frac{{\sinh \left( {p_B E/\left(k_BT\right)} \right)}}{{\left( {p_B E/\left(k_BT\right)} \right)}}}}{D},
\label{eq:23}
\end{equation}
where
\begin{eqnarray}
D=n_{ + b} V_ +  e^{\left( { - hV_ +   - e\psi } \right)/k_B T}  + n_{ - b} V_ -  e^{\left( { - hV_ -   + e\psi } \right)/k_B T}  + n_{wb} V_w e^{ - hV_w /k_B T} \frac{{\sinh \left( {p_w  E/\left(k_BT\right)} \right)}}{{\left( {p_w E/\left(k_BT\right)} \right)}}+  \nonumber \\  n_{Bb} V_B e^{ - hV_B /k_B T} \frac{{\sinh \left( {p_B E/\left(k_BT\right)} \right)}}{{\left( {p_B E/\left(k_BT\right)} \right)}}\nonumber.
\end{eqnarray}.
It should be noted that for the case without Bjerrum pair, our approach corresponds to one of \cite%
{iglic_1996,  iglic_2010, iglic_2015, sin_2016}. 
Substituting Eq. (\ref{eq:20},\ref{eq:21},\ref{eq:22},\ref{eq:23}) in Eq. (\ref{eq:19}), the following equation is obtained
\begin{eqnarray}
n_{ + b} \left( {e^{\left( { - hV_ +   - e\psi } \right)/k_B T}  - 1} \right) + n_{ - b} \left( {e^{\left( { - hV_ -   + e\psi } \right)/k_B T}  - 1} \right) + \nonumber \\ n_{wb} \left( {e^{ - hV_w /k_B T} \frac{{\sinh \left( {p_w E/\left(k_BT\right)} \right)}}{{\left( {p_w E/\left(k_BT\right)} \right)}} - 1} \right) + n_{Bb} \left( {e^{ - hV_B /k_B T} \frac{{\sinh \left( {p_B E/\left(k_BT\right)} \right)}}{{\left( {p_B E/\left(k_BT\right)} \right)}} - 1} \right) = 0.
\label{eq:24}
\end{eqnarray}
Performing minimization of Lagrangian with respect to $\psi \left( r \right)$, the following equation is given 
\begin{equation}
\frac{{\delta L}}{{\delta \psi }} = \frac{{\partial ^2 L}}{{\partial r\left( {\partial \left( {\nabla \psi } \right)} \right)}} - \frac{{\partial L}}{{\partial \psi }} = 0.
\label{eq:25}
\end{equation}
Therefore, we obtain the Poisson equation that determines electrostatic potential, i.e
\begin{equation}
\nabla \left( {\varepsilon _0 \varepsilon _{r} \nabla \psi } \right) =  - e\left( {n_ +   - n_ -  } \right),
\label{eq:26}
\end{equation}
where effective relative permittivity is given by the following equation
\begin{eqnarray}
 \varepsilon _{r}  = 1 + \frac{P}{{\varepsilon _0 E}}  = 1+ \frac{{n_w p_w L\left( {p_w E/\left(k_BT\right)} \right)}}{{\varepsilon _0 E}} + \frac{{n_B p_B L\left( {p_B E/\left(k_BT\right)} \right)}}{{\varepsilon _0 E}},
\label{eq:27}
\end{eqnarray}
where $L\left(x\right)=coth\left(x\right)-\frac{1}{x}$ is the Langevin function.

The following boundary conditions for solving the Poisson equation are used
\begin{eqnarray}
\psi \left( {x \to \infty } \right) = 0,
E\left( {x = 0} \right) =  - \frac{\sigma }{{\varepsilon _0 \varepsilon_{r}\left(x=0\right) }},
\label{eq:28}
\end{eqnarray}
where $\sigma$ is the surface charge density of the charged surface.
So far we discussed the electrostatic properties of the electrolyte near a charged plate. 
Now, let's study for the case of two parallel charged surfaces. 
The authors of \cite%
{yaakov_2009} proved that for the case when free energy density does not explicitly depend on the spatial variables, the osmotic pressure between two charged surfaces is determined by the following formula 
\begin{equation}
f - \left( {\partial f/\partial \psi '} \right)\psi ' = const =  - P,
\label{eq:29}
\end{equation}
where $P$ is the local pressure that is the sum of the osmotic pressure $\Pi$ and the bulk pressure $P_{bulk}$.
Here, we emphasize that the alternative way to derive Eq. (\ref{eq:30}) is to integrate  Eq. (\ref{eq:26}), as demonstrated in \cite%
 {misra_2013, gongadze_2014, gongadze_2013,verwey_1948}. 

After substituting Eq. (\ref{eq:1}) in Eq. (\ref{eq:29}), we get
\begin{equation}
P =  - \frac{{\varepsilon _0 E^2 }}{2} - e_0 z\psi \left( {n_ +   - n_ -  } \right) + \mu _ +  n_ +   + \mu _ -  n_ -   + \left\langle {\mu _w \left( \omega  \right)\rho _w \left( \omega  \right)} \right\rangle   + \left\langle {\mu _B \left( \omega  \right)\rho _B \left( \omega  \right)} \right\rangle   + TS.
\label{eq:30}
\end{equation}
Considering the fact that as $H\rightarrow \infty$, $P\left( {H = \infty } \right) = P_{bulk}$, we get the following equation	
\begin{equation}
P_{bulk}  = \lambda _b  = \left( {2n_b  + n_{wb}  + n_{Bb} } \right)k_B T.
\label{eq:31}
\end{equation}
Comparing Eq.(\ref{eq:30}) and Eq.(\ref{eq:31}), we get the following mathematical expression
\begin{equation}
\Pi  =  - \frac{{\varepsilon _0 E^2 }}{2} + k_B Th - k_B Tn_w \left( {p_w E/\left(k_BT\right)} \right)L\left( {p_w E/\left(k_BT\right)} \right) - k_B Tn_B \left( {p_B E/\left(k_BT\right)} \right)L\left( {p_B E/\left(k_BT\right)} \right).
\label{eq:32}
\end{equation}

 If we neglect the difference in size between different ions and solvent dipoles and the formation of Bjerrum pair, Eq.(\ref{eq:32}) is reduced to the same formula as in \cite%
{misra_2013, gongadze_2014}. 

Taking into account Eq. (\ref{eq:31}) provides a modified expression of dissociation-association reaction.
 \begin{eqnarray}
K=\frac{{\left( {n_b  - n_{Bb} } \right)^2 }}{{n_{Bb} }} \simeq \frac{1}{V_w} \exp \left(  - \frac{\left( {V_ +   + V_ -   - V_B } \right)}{V_w} \right)\exp \left( { - J/k_B T} \right).
\label{eq:33}
\end{eqnarray}

It should be emphasized that our approach has an important advantage as compared to the method of \cite%
{adar_2017}.
The approach can account for not only different sizes of ions, solvent molecules and Bjerrum pairs but also different values of a Bjerrum dipole moment. 

In fact, the approach of \cite%
{adar_2017} is based on the assumption that ions, solvent molecules and Bjerrum pairs have an equal size. As a result, the lattice parameter does not denote the effective size of water dipoles. Moreover, the dipole moment of a solvent molecule has an unreasonable value(9.8$D$)  much higher than the realistic value.

However, unlike in \cite%
{adar_2017}, our approach can use the effective volumes of ions, Bjerrum pairs and solvent molecules and the reasonable values of dipole moment of a solvent molecule and a Bjerrum pair.

\section{Results and Discussion}

Taking into account steric effects and solvent polarization, we first consider the variations in counterion number density, Bjerrum pair number density, water molecule number density and relative permittivity.
\begin{figure}
\includegraphics[width=1\textwidth]{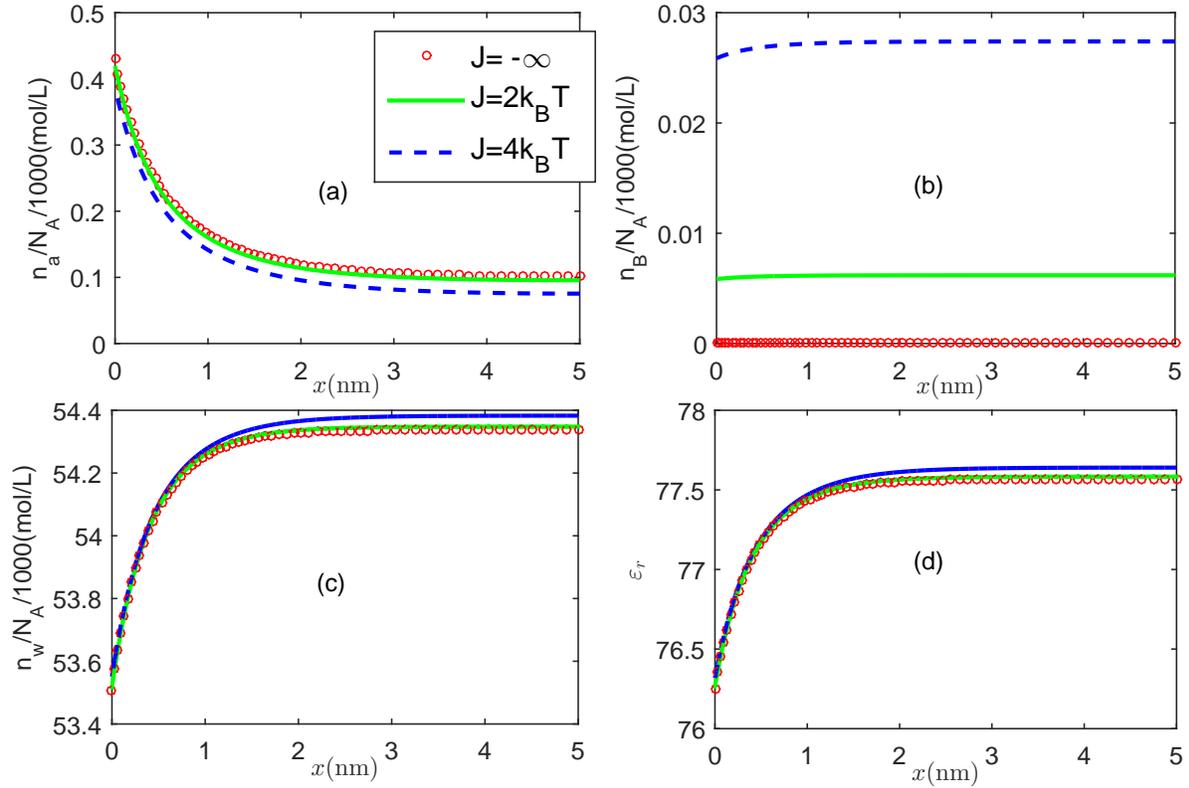}
\caption{ (Color online) The counterion number density (a),  Bjerrum pair number density (b),  water molecule number density (c) and relative permittivity (d) as a function of the distance near a charged surface for the cases with different values of ion-pair association energy ($J=-\infty, 2k_BT, 4k_BT$). Circles, solid line and dashed line denote $J=-\infty, 2k_{B}T$ and $4k_{B}T$, respectively. 
Here solvent is water which has  the relative permittivity of 78.5,  molecular weight of 18g/mol,  density of mass of 1000$kg/m^3$ and temperature is T=300K, the bulk ion cocentration $c_{b}=0.1$mol/L, the volume of an anion $V_-=0.10nm^3$, the volume of an cation $V_+=0.10nm^3$, the volume of a Bjerrum pair $V_B=0.15nm^3$, the surface charge density $\sigma=0.03C/m^{2}$, the dipole moment of a Bjerrum  pair $p_{B}=0.5p_{w}$.}
\label{fig:2}
\end{figure}

In order to set the same dissociation constants as in \cite%
{adar_2017}, we use $\varepsilon_p =78,  V_+=V_-=0.1nm^3, V_B=0.15nm^3, T=300K$ for aqueous solution, while $\varepsilon_p =20,  V_+=V_-=0.1nm^3, V_B=0.18nm^3, T=300K$ for ethylalcohol electrolyte solution.
The reason for the volumes can be understood by Eq. (\ref{eq:33}).

Applying the formula of permittivity Eq. (\ref{eq:27}) to the case of bulk solutions, it is derived that the dipole moment of a water molecule is 4.8D and the dipole moment of a ethylalcohol molecule is 4.26D, where 1D=$3.336\times 10^{-30}C\cdot m$.

In references \cite%
{roij_2009, adar_2017}, it was widely recognized that in aqueous electrolyte solutions, the dipole moment of a Bjerrum pair has only a lower values than ones of water dipoles, whereas in alcohol electrolyte solution, it may be higher than one of alcohol dipoles.
\begin{figure}
\includegraphics[width=1\textwidth]{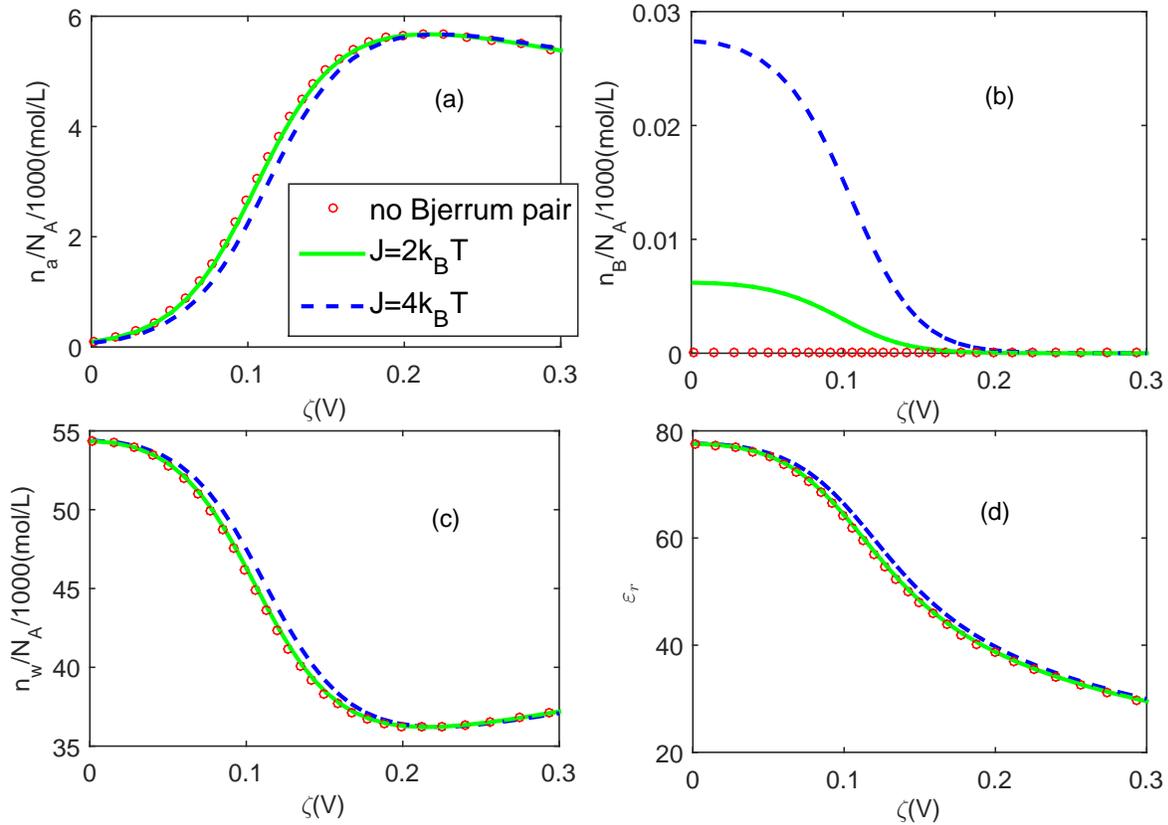}
\caption{(Color online) The counterion number density (a),  Bjerrum pair number density (b),  water molecule number density (c) and relative permittivity (d) as a function of the surface voltage for the cases with the different values of ion-pair association energy ($J=-\infty, 2k_BT, 4k_BT$). Circles, solid line and dashed line denote $J=-\infty, 2k_{B}T$ and $4k_{B}T$, respectively. Other parameters are the same as in Fig. \ref{fig:2}.}
\label{fig:3}
\end{figure}

Fig. \ref{fig:2}(a) depicts the spatial variation in the counterion number density for different values of $J$($J=-\infty, 2k_BT, 4k_BT$). Here $c_b=0.10$mol/L, $V_-=0.10nm^3,V_+=0.10nm^3,V_p=0.15nm^3,\sigma=0.03C/m^2, p_B=0.5p_w$.
Physically, a higher value of $J$ signifies a larger decrease of bulk electrolyte ion concentration, leading to corresponding increase of Bjerrum pair number density. It is clearly seen that as the distance is decreased, the influence of ion association on the counterions distribution is weakened, and therefore, the counterion curves show very little deviation from one another. 
 
Fig. \ref{fig:2}(b) depicts the Bjerrum pair number density as a function of the distance from a charged surface.
In this case, Bjerrum pair number density is hardly changed with the distance from the electrode. 
This is attributed to the fact that in the electric field made by a low surface charge density, a Bjerrum pair with a low dipole moment is only weakly forced.  

Fig. \ref{fig:2}(c) depicts the water molecule number density as a function of the distance from a charged surface.
As we can expect, water dipole moments are depleted near a charged surface. In addition, it is seen that the lowering of counterion number density due to ion association increases water molecule number density, according to the incompressibility condition. 

Fig. \ref{fig:2}(d) depicts the relative permittivity as a function of the distance from a charged surface.
Here we first note that when in the presence of Bjerrum pair, bulk relative permittivity is high compared to the case without Bjerrum pairs.  According to the permittivity formula of Eq. (\ref{eq:27}), this is attributed to the fact that electrolyte permittivity is proportional to water molecule number density and Bjerrum pair number density.
As a result, the permittivity in the presence of Bjerrum pair is higher than one in the absence of Bjerrum pair.
 
In order to get more complete understanding of ion association on electrostatic properties, it is necessary to consider electrostatic properties at the charged surface with surface potential. 

Fig. \ref{fig:3}(a) depicts the counterion number density as a function of the surface potential for different values of ion-pair association energy, ($J=-\infty, 2k_BT, 4k_BT$).
For all the cases, as pointed out in \cite%
{sin_2016}, a counterion number density curve first increases with surface potential, reach  a maximal value and then decreases. In fact, this non-monotonic behavior is attributed to steric effects of ions and water molecules.
Here, we focus on the effect of ion association on electrostatic properties.
 From the figure, it is seen that in the region of low surface potential ($ <0.2V$), ion association lowers counterion number densities at the charged surface. 
However, at high surface potentials, the difference in counterion number density at the charged surface for different strengthes of ion association is diminished by counterion saturation close to the charged surface. 

\begin{figure}
\includegraphics[width=0.6\textwidth]{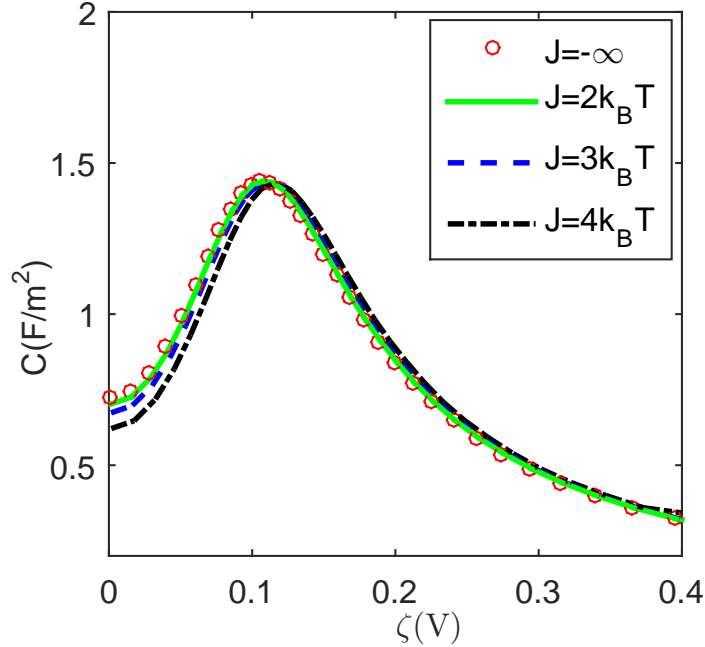}
\caption{(Color online) The differential capacitance as a function of the surface voltage for the cases with the different values of ion-pair association energy ($J=-\infty, 2k_BT, 3k_BT, 4k_BT$). Circles, solid line, dashed line and dish-dotted line denote $J=-\infty, 2k_{B}T, 3k_{B}T $ and $4k_{B}T$, respectively. Other parameters are the same as in Fig. \ref{fig:2}.}
\label{fig:4}
\end{figure}

Fig. \ref{fig:3}(b) depicts the Bjerrum pair number density at a charged surface as a function of the surface potential for three values of  
$J$($J=-\infty, 2k_BT, 4k_BT$). In the presence of Bjerrum pairs, the density is decreased with increasing the surface potential and nearly approaches zero. An increase in the surface potential leads to an enhanced competition between water dipoles and Bjerrum pairs, providing a larger amount of water dipoles. As a result, as the surface potential is increased, Bjerrum pair number density is excluded from the closest proximity of the charged surface.

Fig. \ref{fig:3}(c) depicts the water molecule number density at a charged surface as a function of the surface potential for three different values of ion-pair association energy ($J=-\infty, 2k_BT,4k_BT$). Ion association lowers bulk counterion number density, resulting in an increase of water molecule number density at the charged surface. In particular, it is seen that at low surface potentials($<0.2V$), the difference in water molecule number density between different cases is clearly exhibited, whereas in the higher region of surface potential($>0.2V$), the difference disappears. 

Fig. \ref{fig:3}(d) depicts the relative permittivity at a charged surface as a function of the surface potential for three different values of ion-pair association energy($J=-\infty, 2k_BT, 4k_BT$). Ion association results in an increase of relative permittivity at the charged surface. The reason for this enhancement is that as in Fig. \ref{fig:3}(c), the presence of Bjerrum pairs leads to an increase in water molecule number density. At low surface potentials($<0.2V$), the difference in permittivity between different cases is clearly exhibited but at the higher values of surface potential($>0.2V$) disappear. 

\begin{figure}
\includegraphics[width=1\textwidth]{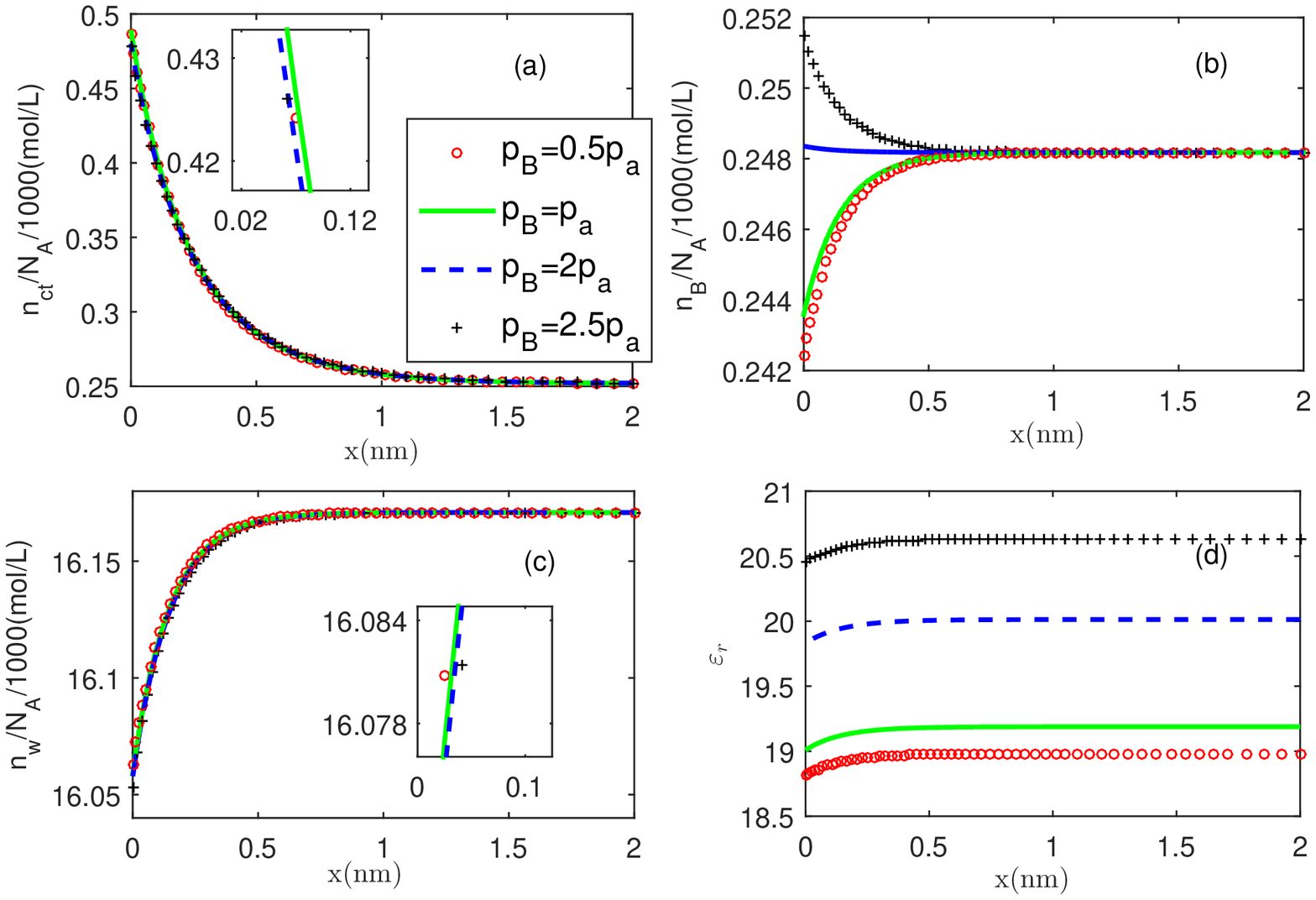}
\caption{(Color online) (Color online) The counterion number density (a),  Bjerrum pair number density (b),  water molecule number density (c) and relative permittivity (d) as a function of the distance from a charged surface for the cases with different values of a Bjerrum dipole moment ($ p_{B}=0.5p_{a}, p_{a}, 2p_{a}, 2.5p_{a} $). Circles, solid line, dashed line, plus signs denote $ p_{B}=0.5p_{a}, p_{a}, 2p_{a}, 2.5p_{a} $, respectively. 
Here the solvent is ethylalcohol which has  the relative permittivity of 20,  molecular weight of 46g/mol, density of mass of 789$kg/m^3$ and $ T=300K, c_{b}=0.5mol/L, J=4k_BT, V_-=0.10nm^3, V_+=0.1nm^3, V_B=0.18nm^3, \sigma=0.01C/m^{2}$.}
\label{fig:5}
\end{figure}
 
Fig. \ref{fig:4} depicts the differential capacitance as a function of surface potential. All the differential capacitances are non-monotonic functions of surface potential and show the same behavior. They first increase at low voltages, then have maxima at intermediate voltages and slowly decrease toward zero at higher voltages. 
It is clearly seen that ion association lowers differential capacitance.
Because the bulk ion number density($c_b=0.1$mol/L) is not so high, Bjerrum pair number density is also low and consequently  the differential capacitance is affected only by the decrease in bulk electrolyte concentration, but not the effect due to polarization of Bjerrum pairs.
 
Fig. \ref{fig:5}(a) depicts the counterion number density as a function of the distance from the charged surface for different values of a Bjerrum pair dipole moment in alcohol electrolyte solution for $\sigma=+0.01C/m^2$. 
It is seen that counterion number densities for all the cases are equal for low surface charge densities. This is explained by the fact that the low surface charge density induces a weak electric field, so that there does not exist the variation due to difference in dipole moments of Bjerrum pairs 

Fig. \ref{fig:5}(b) depicts the Bjerrum pair number density as a function of the distance from the charged surface for the cases having different dipole moments of a Bjerrum pair. First of all, it should be noted that for the low values of Bjerrum-pair dipole moment($p_B=0.5p_w,p_w$), the Bjerrum pair number density decreases with decreasing the distance from the charged surface, whereas for the high values($p_B=2p_w,2.5p_w$), the Bjerrum pair number density increases with decreasing the distance. In the same way as in \cite%
{sin_pccp_2016}, this is a consequence of the competition effect of water dipoles and Bjerrum pairs to occupy near a charged surface. 

\begin{figure}
\includegraphics[width=1\textwidth]{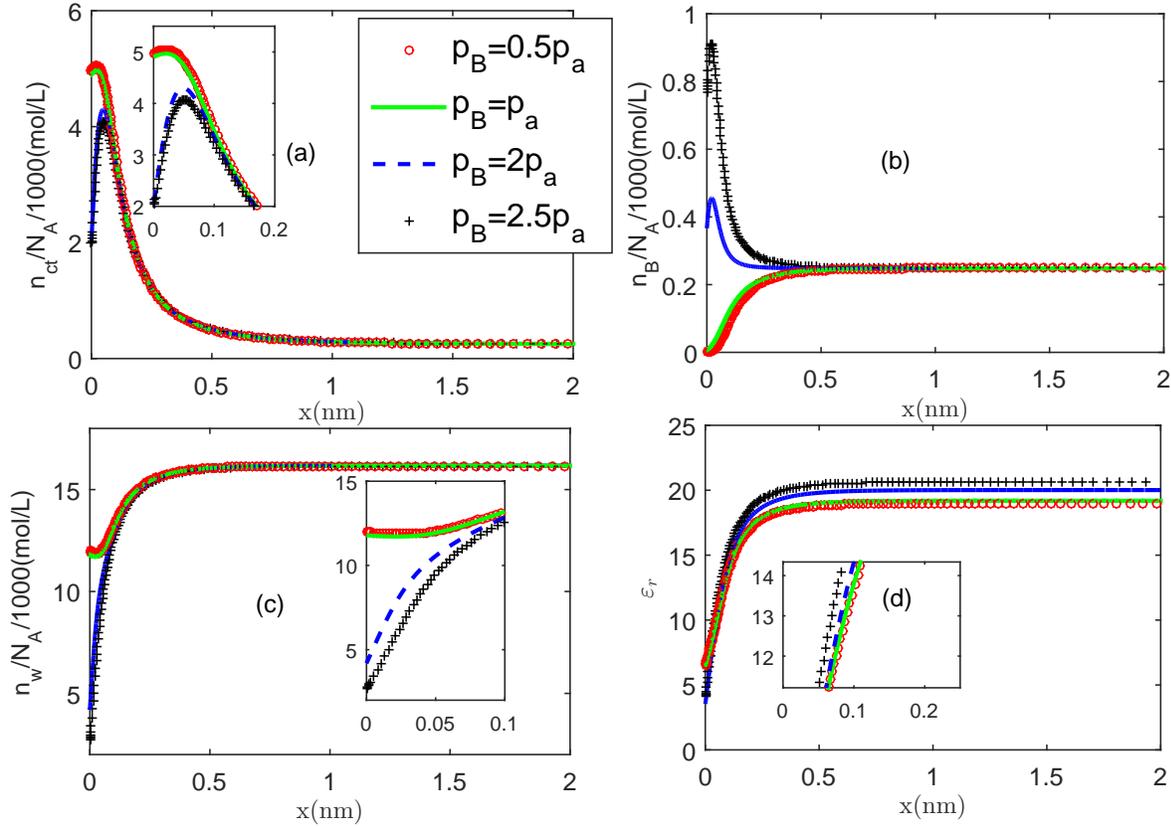}
\caption{(Color online) The counterion number density (a),  Bjerrum pair number density (b),  water molecule number density (c) and relative permittivity (d) as a function of the distance near a charged surface for the cases with different values of a Bjerrum dipole moment ($ p_{B}=0.5p_{a}, p_{a}, 2p_{a}, 2.5p_{a} $). Circles, solid line, dashed line, plus signs denote $ p_{B}=0.5p_{a}, p_{a}, 2p_{a}, 2.5p_{a} $, respectively. 
Here  the surface charge density is $\sigma=0.1C/m^2$ and other parameters are the same as in Fig. \ref{fig:5}.}
\label{fig:6}
\end{figure}

\begin{figure}
\includegraphics[width=1\textwidth]{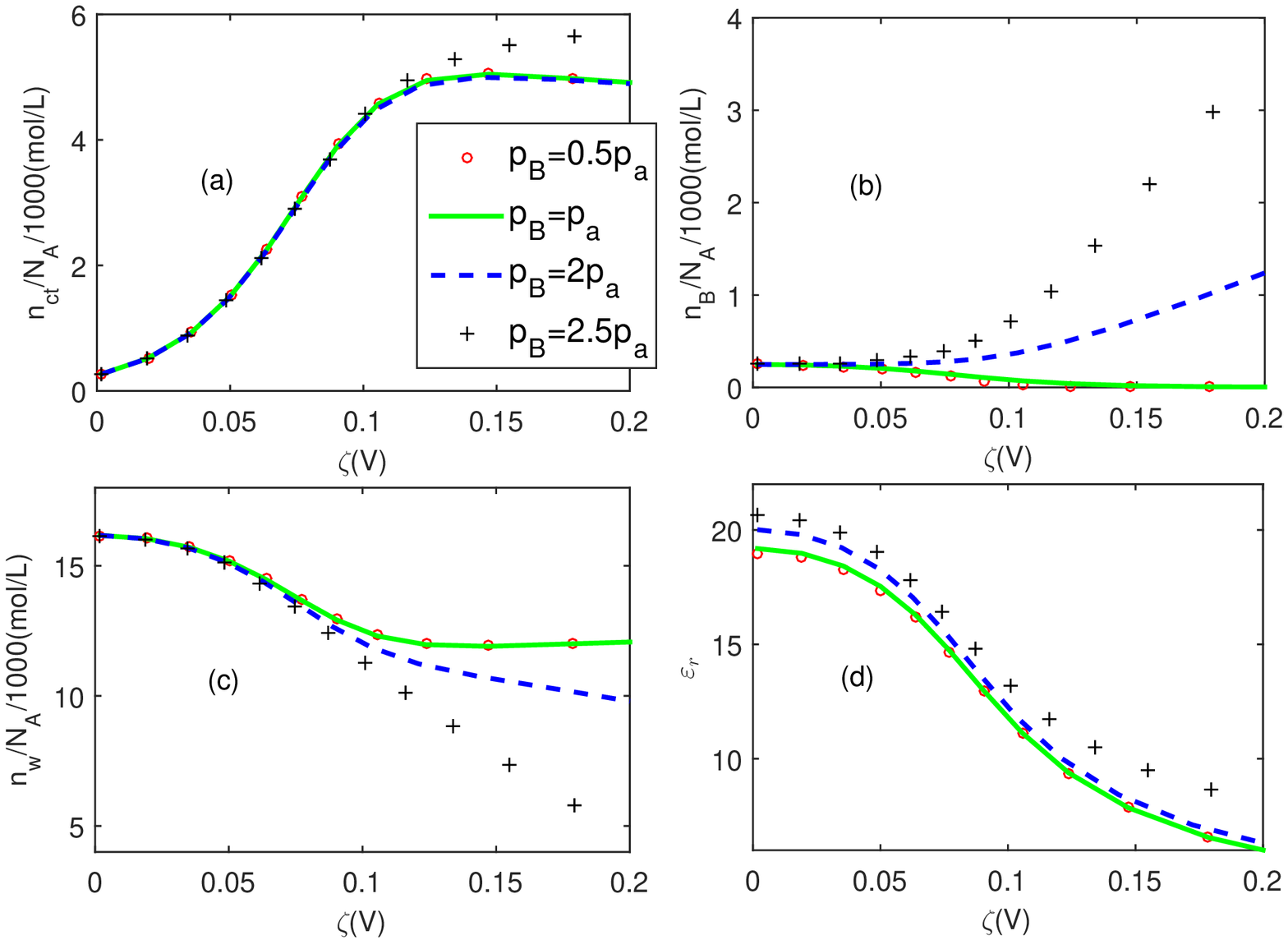}
\caption{(Color online)  The counterion number density (a),  Bjerrum pair number density (b),  water molecule number density (c) and relative permittivity (d) as a function of the surface voltage fo the cases with different values of a Bjerrum dipole moment ($ p_{B}=0.5p_{a}, p_{a}, 2p_{a}, 2.5p_{a}$). Circles, solid line, dashed line, plus signs denote $ p_{B}=0.5p_{a}, p_{a}, 2p_{a}, 2.5p_{a}$, respectively.  Other parameters are the same as in Fig. \ref{fig:5}.} 
\label{fig:7}
\end{figure}

\begin{figure}
\includegraphics[width=0.6\textwidth]{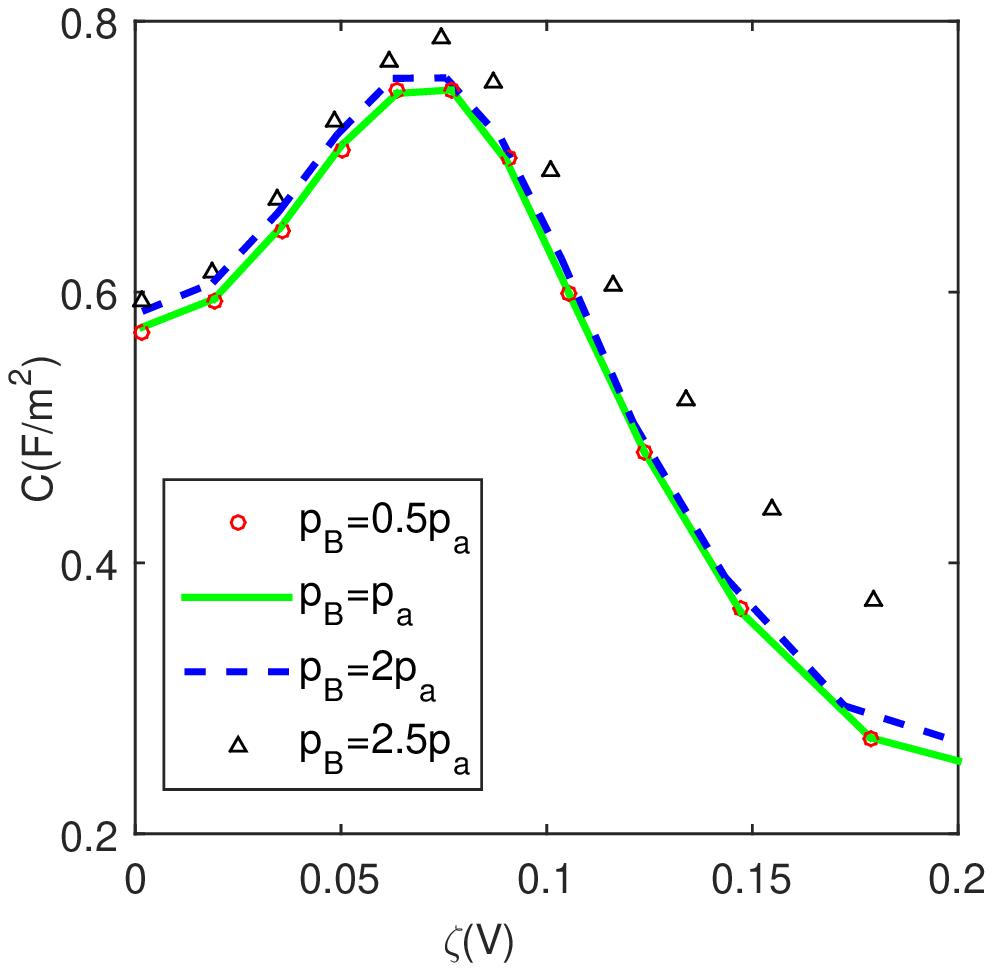}
\caption{(Color online) The differential capacitance as a function of the surface voltage for the cases with different values of a Bjerrum dipole moment ($ p_{B}=0.5p_{a}, p_{a}, 2p_{a}, 2.5p_{a} $). Circles, solid line, dashed line, plus signs denote $ p_{B}=0.5p_{a}, p_{a}, 2p_{a}, 2.5p_{a} $, respectively. Other parameters are the same as in Fig. \ref{fig:5}.
 }
\label{fig:8}
\end{figure}

Fig. \ref{fig:5}(c) depicts the water molecule number density as a function of the distance from the charged surface for different dipole moments of a Bjerrum pair.
A low surface charge density induces a negligible difference in the number density of Bjerrum pairs.
Fig. \ref{fig:5}(d) depicts the relative permittivity of electrolyte solution near a charged surface as a function of the distance from the charged surface for different dipole moments of a Bjerrum pair.
This shows a clear trend in which the relative permittivity of electrolyte solution decreases as the dipole moment of a Bjerrum pair increases. This is explained by the permittivity formula of the Eq.(\ref{eq:27}).
 
Fig. \ref{fig:6}(a)-(d) depict counterion number density, Bjerrum pair number density and alcohol molecule number density and relative permittivity as a function of the distance from the charged surface for different dipole moments of a Bjerrum pair in alcohol electrolyte solution for $\sigma=+0.1C/m^2$, respectively. 

The important thing is that in the case having a high value of Bjerrum pair dipole moment, the counterion number density at the charged surface is lower than corresponding ones for lower values. This is understood by comparing Fig. \ref{fig:6}(b) and Fig. \ref{fig:6}(c). In fact, the dipoles with different sizes and different dipole moments compete each other to occupy locations close to the charged wall. In \cite%
{sin_pccp_2016}, it was confirmed that the ratio of dipole moment to dipole size is the key factor in the competition. As shown in Fig. \ref{fig:6}(b) and Fig . \ref{fig:6}(c), in the cases where $p=2.5p_w$, the alcohol molecule number density decreases from 12mol/L to 5mol/L, whereas Bjerrum pair number density increases from 0.3mol/L to 0.8mol/L. Then, counterion number density at the charged surface increases according to the the incompressibility condition.

Fig. \ref{fig:6}(d) represents that a higher value of Bjerrum-pair dipole moment results in a higher value of relative permittivity. This is obvious, by combining Bjerrum-pair dipole number density (see Fig. \ref{fig:6}(b)) and permittivity formula Eq. (\ref{eq:27}). 
As alcohol is a low polar solvent, the Debye length shortens compared to aqueous electrolyte solutions. As a consequence, as shown in Fig. \ref{fig:5}(a-d) and \ref{fig:6}(a-d), electrostatic properties in alcohol electrolyte solution is mainly changed only inside the $1nm$ region from the charged surface.
 
Fig. \ref{fig:7}(a) depicts the counterion number density as a function of the surface potential for different values of Bjerrum dipole moment . It can know that in the region of low surface potential ($<0.1$), different values of Bjerrum pair dipole moment($p_B=0.5p_a, p_a, 2p_a,  2.5p_a$) do not provide the difference in counterion number density. 
However, at high surface potentials (0.1V$< \zeta <$0.2V), the counterion number density at the charged surface for a high value of Bjerrum pair dipole moment is higher than ones for a lower value of dipole moment. The reason for this phenomena is that as above-mentioned, there exists the competition between alcohol dipoles and Bjerrum pair dipoles near the charged surface. An increase in surface voltage enhances the difference in counterion number density between the different cases.

Fig. \ref{fig:7}(b) depicts the Bjerrum pair number density at a charged surface as a function of the surface potential for four Bjerrum pair dipole moments.  It is clear that in the region($>$0.1V), for the cases with high values of Bjerrum pair dipole moment, Bjerrum pair number density rapidly increases due to the above-mentioned competition.

Fig. \ref{fig:7}(c) depicts the alcohol molecule number density at a charged surface as a function of the surface potential for four Bjerrum pair dipole moments ($p_B=0.5p_a, p_a, 2p_a,2.5p_a$).  For the cases where $p_B=2p_a, 2.5p_a$, the rapid decrease of alcohol dipole number density at the charged surface is attributed to the effect due to competition between different dipoles.

Fig. \ref{fig:7}(d) depicts the relative permittivity at a charged surface as a function of the surface potential for different Bjerrum pair dipole moments. It should be noted that for the cases with $p_B=0.5p_a, p_a,2 p_a$, the permittivity curves behave in the same way, whereas, the permittivity for the case with $p_B=2.5p_a$ very slowly decreases with the surface voltage. This can be explained as follows:
On one hand, at high potentials, alcohol molecule number density drastically diminishes with the surface potential, whereas Bjerrum pair number density sharply increases with the potential. On the other hand, Bjerrum-pair dipole moment is larger than alcohol dipole moment.
Considering the two facts, Eq. (\ref{eq:27}) can ensure for the permittivity slowly to decrease. 

\begin{figure}
\includegraphics[width=1\textwidth]{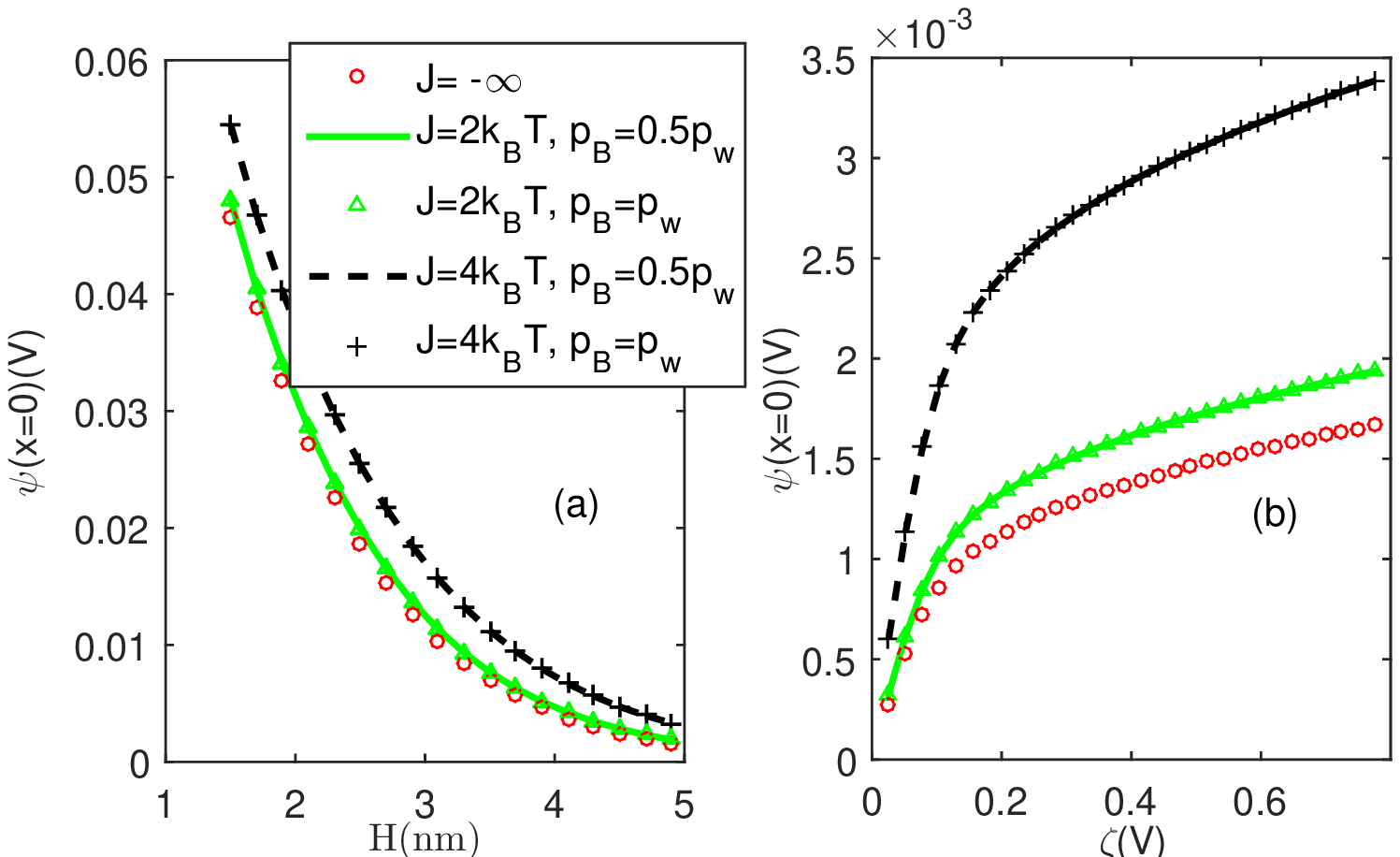}
\caption{(Color online) For similarly charged surfaces, (a) variation of the centerline potential with the separation distance between the charged surfaces for $\psi(x = H/2)= \psi(x = -H/2) = +0.5V$. (b)Variation of the centerline potential with the surface potential for different values of ion-pair association energy and a Bjerrum dipole moment. The separation distance between charged surfaces is $H = 5 nm$. Circles, solid line, triangles, dashed line and plus signs represent the cases having  ($J=-\infty$ without Bjerrum pair), ($J=2k_BT, p_B=0.5p_w$), ($J=2k_BT, p_B=p_w$), ($J=4k_BT, p_B=0.5p_w$), ($J=4k_BT, p_B=p_w$), respectively. The solvent is water and other parameters are the same as in Fig. \ref{fig:2}.}
\label{fig:9}
\end{figure}

\begin{figure}
\includegraphics[width=1\textwidth]{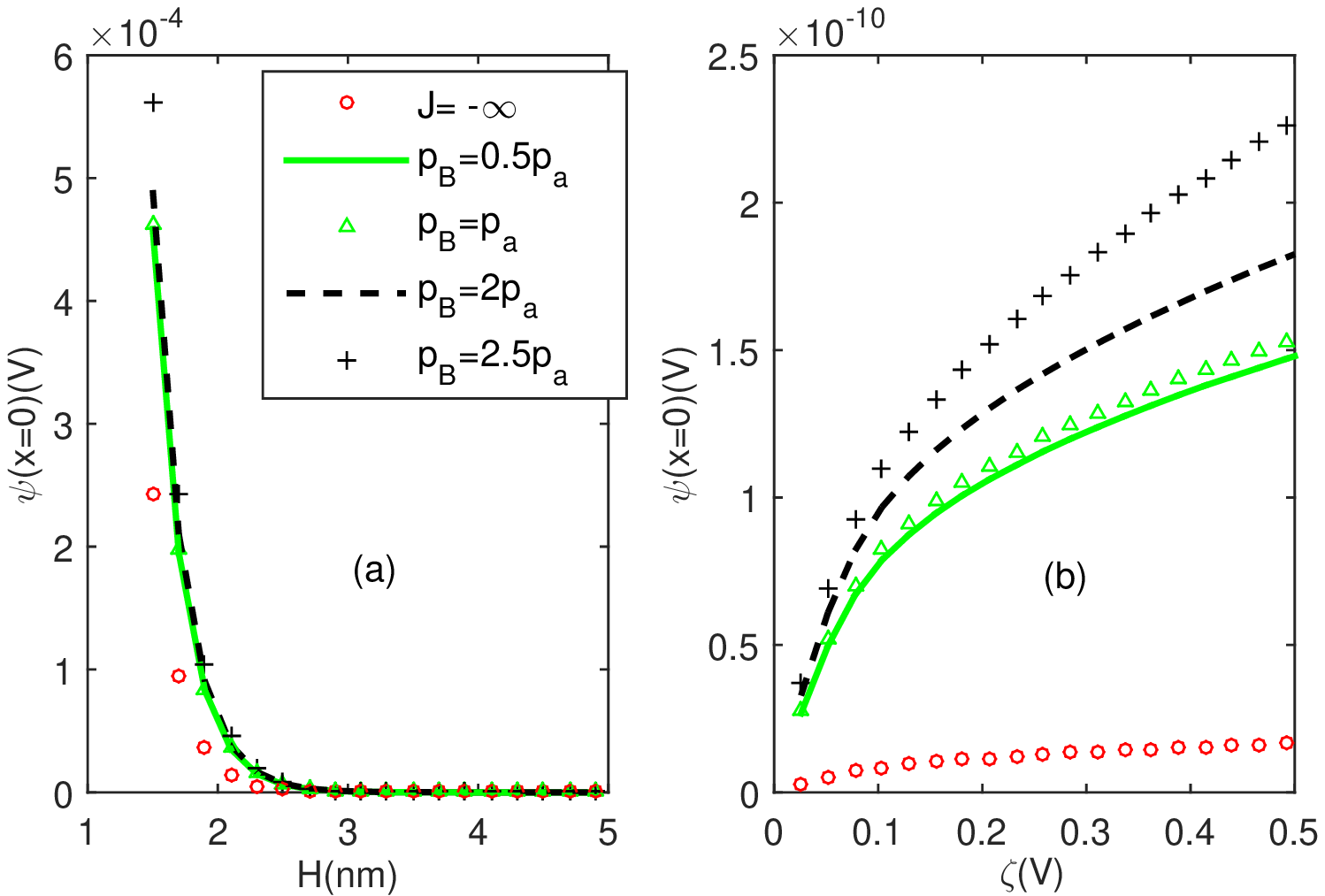}
\caption{(Color online) For similarly charged surfaces, (a) variation of the centerline potential with the separation distance between the charged surfaces for $\psi(x = H/2)= \psi(x = -H/2) = +0.5V$. (b)Variation of the centerline potential with the surface potential for different values of ion-pair association energy and a Bjerrum dipole moment. The separation distance between charged surfaces is $H = 2 nm$. Circles, solid line, triangles, dashed line and plus signs represent the cases having  ($J=-\infty$ without Bjerrum pair), ($J=2, p_B=0.5p_a$), ($J=2, p_B=p_a$), ($J=2, p_B=2p_a$), ($J=2, p_B=2.5p_a$), respectively. The solvent is ethylalcohol and other parameters are the same as in Fig. \ref{fig:5}.}
\label{fig:10}
\end{figure}

\begin{figure}
\includegraphics[width=1\textwidth]{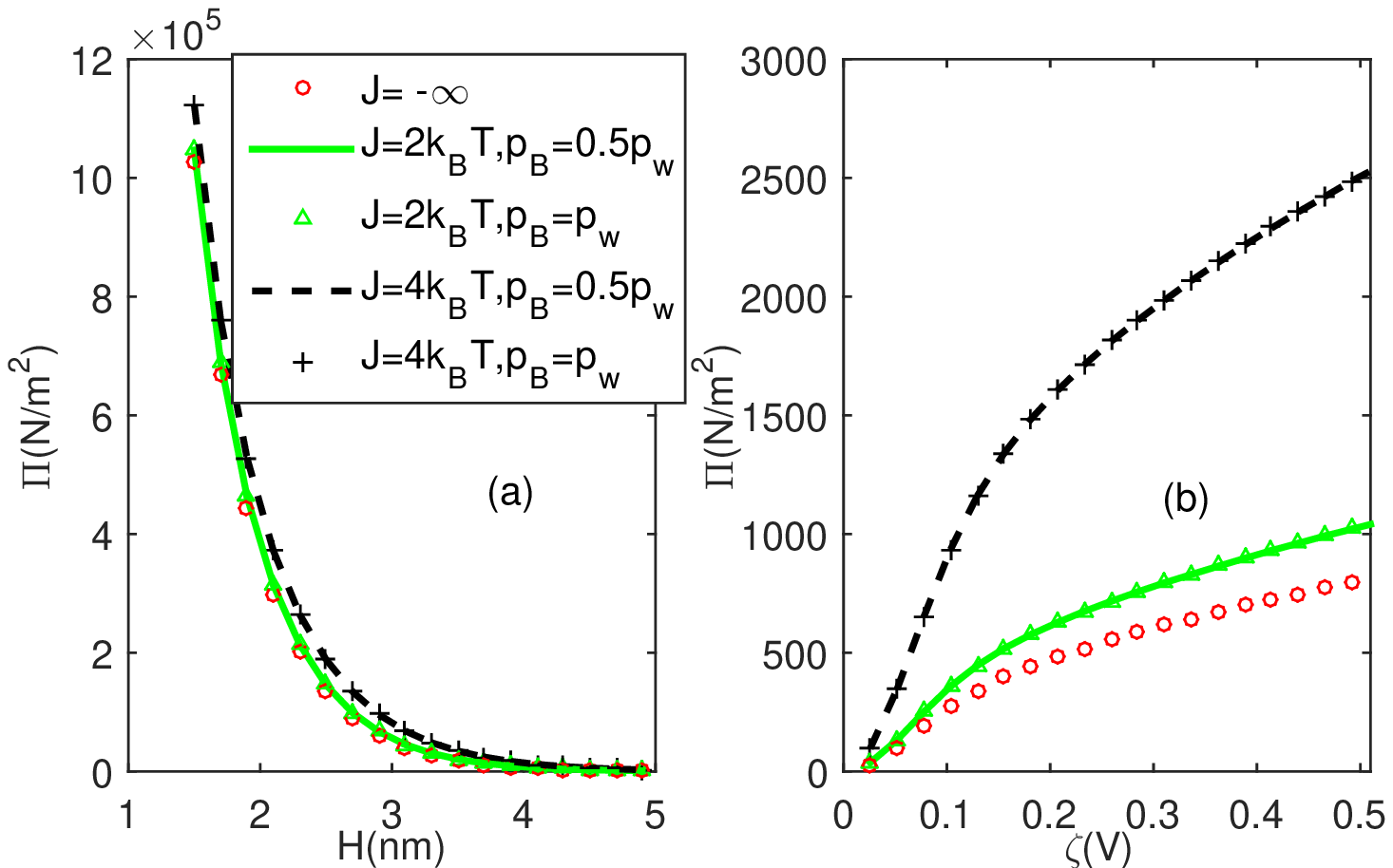}
\caption{(Color online) For similarly charged surfaces, (a) variation of the osmotic pressurel with the separation distance between the charged surfaces for $\psi(x = H/2)= \psi(x = -H/2) = +0.5V$. (b)Variation of the osmotic pressure with the surface potential for different values of ion-pair association energy and a Bjerrum dipole moment. The separation distance between charged surfaces is $H = 5 nm$. Circles, solid line, triangles, dashed line and plus signs represent the cases having  ($J=-\infty$ without Bjerrum pair), ($J=2k_BT, p_B=0.5p_w$), ($J=2k_BT, p_B=p_w$), ($J=4k_BT, p_B=0.5p_w$), ($J=4k_BT, p_B=p_w$), respectively. The solvent is water and other parameters are the same as in Fig. \ref{fig:2}.}
\label{fig:11}
\end{figure}

\begin{figure}
\includegraphics[width=1\textwidth]{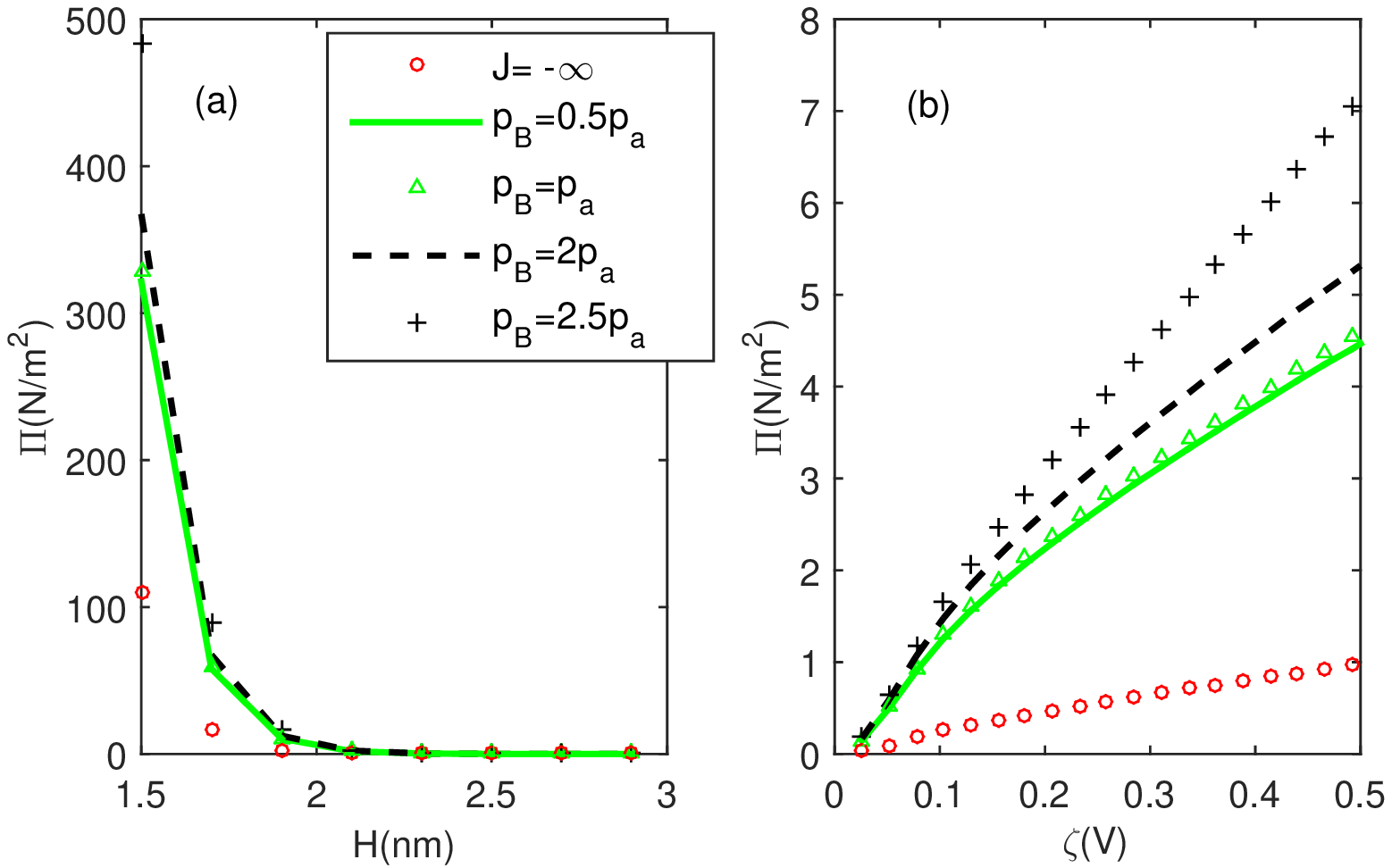}
\caption{(Color online)For similarly charged surfaces, (a) variation of the centerline potential with the separation distance between the charged surfaces for $\psi(x = H/2)= \psi(x = -H/2) = +0.5V$. (b)Variation of the centerline potential with the surface potential for different values of ion-pair association energy and a Bjerrum dipole moment. The separation distance between charged surfaces is $H = 2 nm$. Circles, solid line, triangles, dashed line and plus signs represent the cases having  ($J=-\infty$ without Bjerrum pair), ($J=2k_BT, p_B=0.5p_a$), ($J=2k_B, p_B=p_a$), ($J=2k_BT, p_B=2p_a$), ($J=2k_BT, p_B=2.5p_a$), respectively. The solvent is ethylalcohol and other parameters are the same as in Fig. \ref{fig:5}.}
\label{fig:12}
\end{figure} 

\begin{figure}
\includegraphics[width=1\textwidth]{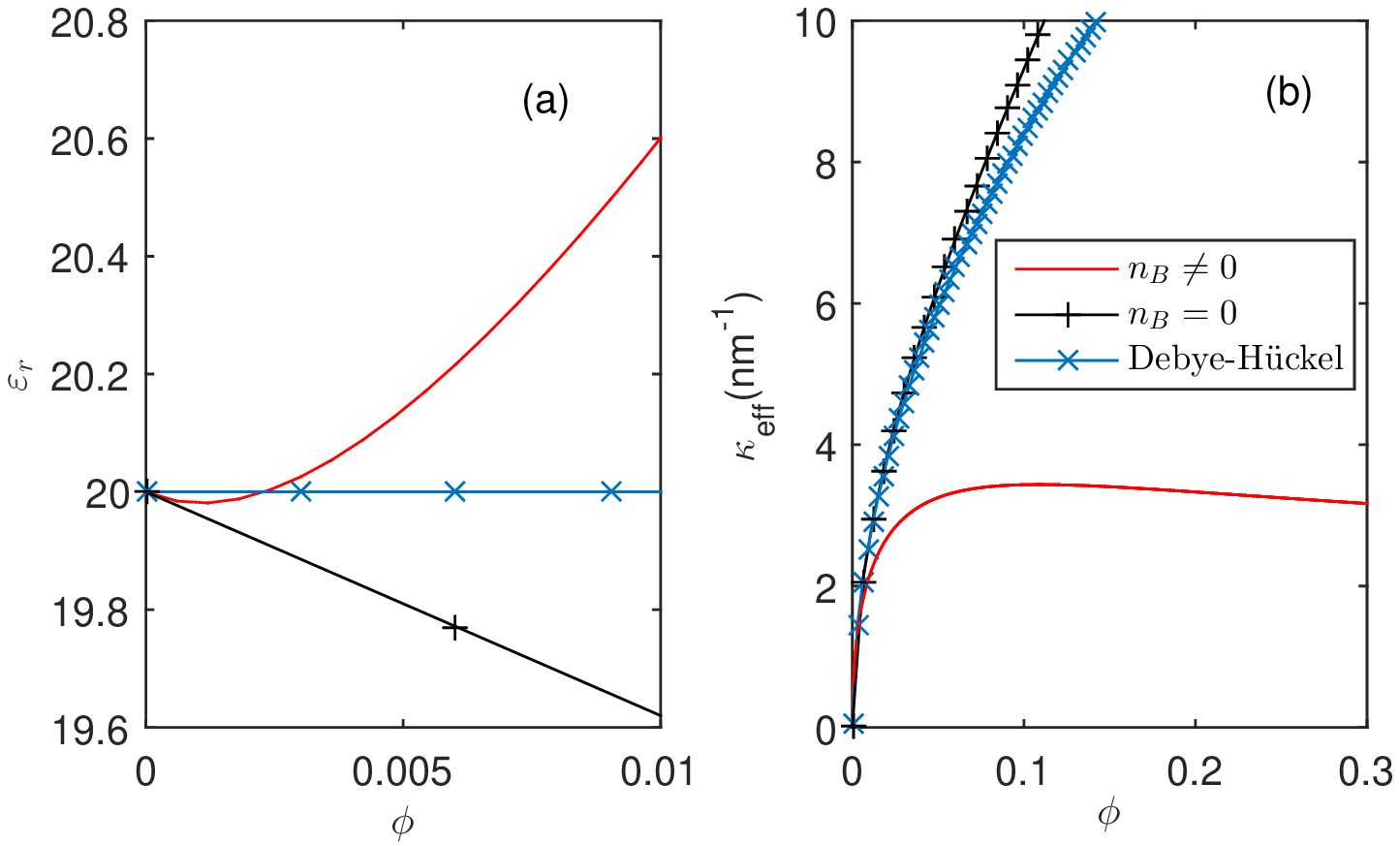}
\caption{(Color online)(a) Relative permittivity, $\varepsilon_r$, and (b) inverse screening length, $\kappa_{eff}$, as a function of the dimensionless bulk ionic concentration $\phi= 0.5n_b \left(V_{-}+V_{+}\right)$  for $p_{B} >  p_{w}$. The results of Eqs. (27) and (34) are shown as a solid red curve and are compared to Eqs. (27) and (34) without Bjerrum pairs ($n_B = 0$, a plus black signs) and to the classical Debye-H\"uckel theory (a cross blue line). The curves are plotted for $T = 298 K$, $J = 4k_BT$, $V_+=V_-=0.1nm^{3}$, $V_{B}=0.17nm^{3}$, $p_B$ = 17D, and $\varepsilon_r=20$.}
\label{fig:13}
\end{figure} 

Fig. \ref{fig:8} depicts the differential capacitance as a function of surface potential for the cases where $p_B=0.5p_a, p_a, 2p_a, 2.5p_a$.
It is noted that for a Bjerrum-pair dipole moment, differential capacitance is higher than ones for smaller values of Bjerrum-pair dipole moment.
In particular, at high voltages, differential capacitance for the case where $p_B=2.5p_a$ is much higher than corresponding one for other cases.
In the same way as in Fig. \ref{fig:7}(d), the reason for this behavior is the competition between alcohol molecules and Bjerrum pairs.

Fig. \ref{fig:9}(a) depicts the electrostatic potential at the centerline between two charged surfaces as a function of the separation for different cases with $J=2k_BT,4k_BT$, $p_B=0.5p_w,p_w$. Here, the concept of centerline means the line or plane consisting of the middle points between two charged surfaces. 
Fig. \ref{fig:9}(a) indicates that for all the cases, an increase of the separation causes a decrease in the centerline potential.
It is also shown that an enhancement of ion association increases the centerline potential.
In fact, as ion association is enhanced, the bulk ion concentration is decreased and consequently, the formation of electric double layer near a charged surface requires a higher electric force. As a result, an enhancement of ion association results in an increase of centerline potential.     
It is also seen that at low bulk counterion concentrations, the difference in Bjerrum-pair dipole moment hardly affects the centerline potential.

Fig. \ref{fig:9}(b) depicts the electrostatic potential at the centerline between two charged surfaces as a function of the surface potential for different cases with $J=2k_BT,4k_BT$, $p_B=0.5p_w, p_w$. The centerline potential increases as surface potential is increased, while the difference in centerline potential between different cases is also enhanced with increasing the surface potential. The fact is attributed to lowering of screening effect of electrolyte solution due to decrease in free ion number density.
 
Fig. \ref{fig:10}(a) depicts the electrostatic potential at the centerline between two charged surfaces as a function of the separation for different cases with $J=2k_BT$ and $p_B=0.5p_a, p_a, 2p_a, 2.5p_a$. 
It is shown that an increase of separation between two charged surfaces decreases the electrostatic potential at the centerline, and reduces the difference between the potentials for other cases. In addition, an increase in dipole moment of Bjerrum pair yields an increase in centerline potential. The reason can be explained as follows: As shown in Fig. \ref{fig:8}(d), the formation of Bjerrum pairs enhances the relative permittivity, that is, the more the Bjerrum pair, the higher the permittivity. The higher the value of dipole moment of Bjerrum pair, the higher the permittivity. On the other hand, if the permittivity of electrolyte solution is high, the screening property of solution gets weak.  Therefore, the centerline potential increases as the dipole moment of Bjerrum pair increases.   
Fig. \ref{fig:10}(b) depicts the electrostatic potential at the centerline between two charged surfaces as a function of the surface potential for different cases with $J=2k_BT$ and $p_B=0.5p_a, p_a, 2p_a, 2.5p_a$. It is evident from the figure that an increase of Bjerrum pair dipole moment results in an increase of the centerline potential. The difference in centerline potential between different cases is also enhanced with increasing the surface potential. 
 
Fig. \ref{fig:11}(a) depicts the osmotic pressure between two charged surfaces as a function of the separation for different cases with $J=2k_BT, 4k_BT$ and $p_B=0.5p_w, p_w$, while Fig. \ref{fig:11}(b) depicts the osmotic pressure as a function of the surface potential for these cases. It is evident from the figure that as ion association is enhanced, the osmotic pressure is increased. In fact, osmotic pressure is proportional to the h value at the centerline between charged surfaces. On the other hand, generally, h value monotonically increases with  the electrostatic potential. Considering surface potential - and spatial dependence of the centerline potential  (see Fig. \ref{fig:9}(a) and (b) ), the above mentioned facts are proved.
 
Fig. \ref{fig:12}(a) depicts the osmotic pressure between two charged surfaces as a function of the separation for different cases with $J=2k_BT, p_B=0.5p_a, p_a, 1.5p_a, 2p_a$ while Fig. \ref{fig:12}(b) depicts the osmotic pressure as a function of the surface potential for these cases. It is evident from Fig. \ref{fig:12}(a) and (b) that as Bjerrum pair has a high electric dipole moment, the osmotic pressure increases. In the same way in Fig. \ref{fig:11}(a) and (b), this should be explained by increasing centerline potential due to the increase of dipole moment value. 

The recent experiment of \cite%
{smith_2016} 
elucidated that in the range of $ \phi$ values close to 0.1, the screening length non-monotonically varies with the bulk ionic concentration, while the authors of \cite%
{adar_2017} theoretically demonstrated that the screening length can increase with ionic concentration for $p > p_w$, unlike in classical Poisson-Boltzmann theory.

Fig. \ref{fig:13}(a) and (b) display the relative permittivity and inverse screening length as a function of the dimensionless ion concentration $\phi=0.5n_b\left(V_+ + V_-\right)$  for the cases considering or not  Bjerrum pairs and for Debye-H\"uckel theory.

We use the same formula of inverse screening length as in \cite%
{adar_2017}. 
\begin {equation}
\kappa _{eff}  = \sqrt {\frac{{2 \left(n_b-n_B\right) }}{{\varepsilon _{0}\varepsilon _{r}k_BT}}} e 
\label{eq:34}
\end {equation}
Fig. \ref{fig:13}(a) shows that the relative permittivity decreases with bulk ion concentration, reaches a minimum value and then increases.
The decreasing behavior is attributed to the reduction in number of solvent molecules due to excluded volume effects of ions.
The increase in the relative permittivity is due to enhancement in the formation of Bjerrum pairs by increase of bulk ion concentration.  
Fig. \ref{fig:13}(b) represents the non-monotonic behavior of $\kappa_{eff}$ for the case of $p_{B}>p_{w}$, as suggested in \cite%
{smith_2016, adar_2017}.

 The above facts imply that the present theory is an effective tool for studying Bjerrum pairs in electrolyte solutions.

Although the present theory uses the value of a single water dipole moment  lower than 9.8D of \cite%
{adar_2017}, the value is still higher than  the dipole moment of a water molecule in bulk liquid water (2.4D-2.6D) \cite%
{dill_2003}.
In fact, as shown in \cite%
{gongadze_2012},  this value of a single water molecule dipole moment can be further decreased to 3.1D by taking into account also electronic polarizability and cavity field of water molecules, which gives also more correct Onsager limit for the bulk solution. However, the method can not provide the analytical solution for the osmotic pressure between two charged surfaces. 

In the future, we should find a free energy formulation where the model value of a single water dipole moment can be additionally decreased by taking into account other factors such as correlations between water dipoles.

\section{Conclusion}

In the present study, we have investigated the effects of ion association on the electrostatic properties in the electrical double layer near  charged surfaces by using a mean-field theory taking into account non-uniform size effect of ions, Bjerrum pairs and solvent molecules and orientational ordering of solvent dipoles and Bjerrum pair dipoles.
  
Our approach accounts for not only different sizes of ions, solvent molecules and Bjerrum pairs but also different dipole moments of solvent molecules and Bjerrum pairs.

In order to assess the effect of ion association on the electric double layer, we have studied the variations in the counterion number density, solvent molecule number density, Bjerrum pair number density and relative permittivity with the distance from a charged surface and with the surface potential for different values of Bjerrum pair concentration and Bjerrum-pair dipole moment. 
We have demonstrated that in an aqueous solution, especially at low surface potentials, ion association brings out decrease of counterion number density and increase of Bjerrum pair number density, water molecule number density and relative permittivity. 
We have further demonstrated that in the alcohol electrolyte solution, the increase of dipole moment of a Bjerrum pair provides increase of counterion number density, Bjerrum pair number density and relative permittivity and decrease of alcohol molecule number density, especially at high surface potentials and at the locations close to the charged surface. 

In aqueous solutions, ion association affects differential capacitance by only decreasing bulk ion concentration. However, the dipole moment of Bjerrum pairs does not affect the differential capacitance. In alcohol electrolyte solution, for the case where the dipole moment of a Bjerrum pair is higher than twice alcohol dipole moment, at high surface potentials, the differential capacitance is much higher than for the case having Bjerrum pair with smaller dipole moments. 

We have also unveiled how the ion association plays an intrinsic role in establishing the electrostatic interaction between the charged surfaces. 
We believe that these findings will be significant in developing a more complete theory of electrolyte solution. 
The present study, accordingly, may act as a theoretical tool which exploits the effects of ion association.

\section{Conflict of interest}
There are no conflicts to declare.

\end{document}